\documentstyle[epsfig,aps,prc,eqsecnum]{revtex}

\def\beq{\begin{equation}}
\def\eeq{\end{equation}}
\def\bea{\begin{eqnarray}}
\def\eea{\end{eqnarray}}
\def\ni{\noindent}

\def\cd {\!\cdot\!}

\def\a{\alpha}

\def\g{\gamma}

\def\m{\mu}

\def\O{\Omega}
\def\p{\pi}

\def\bL {\mbox{\boldmath $L$}}

\def\br {\mbox{\boldmath $r$}}

\def\bsig {\mbox{\boldmath $\sigma$}}
\def\btau {\mbox{\boldmath $\tau$}}
\def\bL {\mbox{\boldmath $L$}}

\begin{document}




\title{Relativistic $O(q^4)$ two-pion exchange\\
nucleon-nucleon potential:\\
parametrized version}

\author{
R. Higa$^{1,}$\footnote{Email address: higa@jlab.org},
M. R. Robilotta$^{2,}$\footnote{Email address: robilotta@if.usp.br} and
C. A. da Rocha$^{3,}$\footnote{Email address: crocha40@usjt.br}}

\address{${}^{1}$ Thomas Jefferson National Accelerator Facility,\\
12000 Jefferson Avenue, Newport News, VA 23606, USA}

\address{${}^{2}$ Instituto de F\'{\i}sica, Universidade de S\~{a}o Paulo,\\
Caixa Postal 66318, 05315-970, S\~ao Paulo, S\~ao Paulo, Brazil}

\address{${}^{3}$ N\'ucleo de Pesquisa em Computa\c c\~ao e Engenharia,
Universidade S\~ao Judas Tadeu,\\
Rua Taquari, 546, 03166-000, S\~ao Paulo, S\~ao Paulo, Brazil}

\date{\today}

\maketitle
\abstract{
The chiral two-pion exchange nucleon-nucleon interaction has nowadays 
a rather firm conceptual basis, but depends on low-energy constants, 
which may be extracted from fits to data. In order to facilitate this 
kind of application, we present here a parametrized version of our 
relativistic expansion of this component of the force to $O(q^4)$, 
performed recently.}

\vspace{5mm}


\section{introduction}

Nuclear interactions are strongly dominated by the quarks $u$ and $d$ 
and can be accomodated into a two-flavor version of $QCD$. The masses 
of these quarks are small in the GeV scale and one is not far from the 
massless lagrangian limit, in which $QCD$ is invariant under chiral 
$SU(2)\times SU(2)$ transformations. For this reason, low energy 
hadronic processes can be reliably described by means of effective 
lagrangians that are symmetric under the Poincar\'e and isospin groups 
and incorporate approximate chiral symmetry, realized in the 
Nambu-Goldstone mode.

The one-pion exchange $NN$ potential $(OPEP)$  is simple, has been 
well established long ago, and dominates completely partial waves 
with orbital momentum $L\geq 5$. The two-pion exchange potential 
$(TPEP)$, on the other hand, is rahter complex and has become free of 
important umbiguities only in the 1990s, after the systematic use of 
chiral symmetry in its theoretical description 
\cite{Bira,etal,SP,BRR,KBW,Kai,Nij,Epe,Mach,HR,HRR}. 

Chiral perturbation theory is based on the existence of a 
characteristic scale $q$, set by both pion four-momenta and nucleon 
three-momenta, such that $q<1$ GeV. Due to this technique, nowadays 
one understands rather well the internal hierarchies of the $NN$ 
potential in terms of chiral layers. Leading terms of the chiral 
$TPEP$ are of order $O(q^2)$ and expansions which go up to $O(q^4)$ 
are already available. One of them was produced recently by our group 
\cite{HR,HRR}. We departed from a relativistic lagrangian and 
evaluated the relevant Feynman diagrams covariantly, without resorting 
to heavy baryon approximations. The so obtained $T$-matrix was then 
transformed into a potential, expressed in terms of covariant loop 
integrals and observable parameters. Without loss of generality, one 
may choose these parameters to be either the subthreshold coefficients 
extracted from $\p N$ scattering \cite{H} or the low-energy contants 
$(LECs)$ present in the effective lagrangian. These two possibilities 
are formally equivalent, but suitded to slightly different physical 
purposes. The former choice yields a closed prediction for the 
potential, whereas the latter gives rise to an open theoretical 
structure which may be used to fit $NN$ data. 

Nowadays, a rather pressing issue is to determine the extent that 
nature backs this picture. The comparison of chiral predictions with 
empirical phase shifts is not straightforward, since existing 
theoretical potentials are not reliable for distances smaller than 
1 fm \cite{HRR}. Three complementary possibilities are available for 
overcoming this problem. The most direct one is to use peripheral 
waves and rely on those windows in angular momenta and energies for 
which the centrifugal barrier is effective in suppressing the 
interaction at short distances. When this happens, the Born 
approximation can be used and one does not need to know the 
potential close to the origin. However, this kind of test is not 
very stringent, since peripheral waves are small and 
uncertainties are 
large \cite{BRR,KBW,R04}. Inclusion of more important waves requires 
the use of  dynamical equations with regularized potentials \cite{Epe}. 
This regularization brings necessarily extra parameters into the 
problem which are not constrained by chiral symmetry. One is then 
faced with the problem of disentangling from ones results those 
windows which do indeed test the symmetry. The third alternative, 
already employed by the Nijmegen group \cite{Nij}, is to assume that 
the theoretical potential determines correctly the interaction in a 
spatial window ranging from a radius $R$ onwards and then use it as 
an input in phase shift analises. This procedure has already proved 
to be useful in the case of the $OPEP$, in the elastic regime and for 
waves with $L\geq 5$. Its extension to the $TPEP$ becomes not trivial 
due to the fact that the range of reliability of the theoretical 
potential depends on the chiral order one is working at. 

The research on the $TPEP$ performed in the last decade has set its 
conceptual foundations on a rather solid basis, comparable to that 
of the OPEP in the late sixties. On the other hand, the $TPEP$ 
depends on several $LECs$, which must be extracted from eihter 
$\pi N$ scattering data or direct fits of $NN$ phase shifts. 
In general, this last kind of procedure tends to be computationally 
heavy, for theoretical results are usually given as cumbersome 
expressions. In order to make applications easier, in this work we 
present a parametrized version of our $O(q^4)$ relativistic 
configuration space $TPEP$, which is  numerically accurate for 
distances larger than 1~fm. It is based on the theoretical 
expressions derived in Ref. \cite{HRR} and reproduced in appendix A, 
as functions of the LECs. 


\section{parametrized potential}

The configuration space potential has the isospin structure 

\beq
V(\br)  =  V^+(\br) + \btau^{(1)} \cd \btau^{(2)} \; V^-(\br)\;,
\label{v1}
\eeq  

\ni
with 

\begin{equation}
V^{\pm}(\br ) = V^{\pm}_{C}
+ V^{\pm}_{LS}\,\Omega_{LS}
+ V^{\pm}_{T}\,\Omega_{T}
+ V^{\pm}_{SS}\,\Omega_{SS}
+V^{\pm}_Q\, \Omega_Q \,,
\label{v2}
\end{equation}

\ni
and $\; \O_{LS}= \bL \cd (\bsig^{(1)}+\bsig^{(2)})/2 ,\;\; 
\O_{T}= 3\, \bsig^{(1)}\cd \hat{\br} \; \bsig^{(2)}\cd \hat{\br}
-\bsig^{(1)}\cd \bsig^{(2)},\;\; 
\O_{SS} = \bsig^{(1)}\cd \bsig^{(2)}$. The form of the operator 
$\O_Q$ in configuration space is highly non-local and can be found 
in Ref. \cite{HM}.

In Refs. \cite{HR,HRR} we have presented a $O(q^4)$ relativistic 
expansion of the $TPEP$, which is reproduced, in an alternative form, 
in appendix A. The configuration space potential is written in terms 
of numerical coefficients which multiply dimensionless functions 
arising form the Fourier transforms of Feynman loop integrals. 
The former are combinations of external parameters representing the 
pion and nucleon masses, $\m$ and $m$, respectively, the pion decay 
constant $f_\p$, the axial coupling constant $g_A$, and the LECs 
$c_i$ and $d_i$. The latter are denoted by $Z_i$ and depend on just 
$\m$, $m$, $f_{\pi}$, and $x \equiv \m\,r$. We keep the external 
quantities as free and parametrize the function $Z_i\equiv(F_i, G_i)$ as 

\beq
Z_i=-\frac{\m}{(4\p)^{5/2}}\, \left( \frac{\m}{f_\p} \right)^4 
\, \left[ \sum \g_i^n \, x^n \right] \,\frac{e^{-2x}}{x^2}  \;.
\label{v3}
\eeq

The coefficients $\g_i^n$ corresponding to the various cases are 
given in the tables at the end of this section. This parametrization 
is more than $1\%$ accurate in the range $0.8$ fm 
$\leq r \leq$ $10$ fm. 

Using the definition $\a \equiv \m/m$, the profile functions are 
written as

\begin{eqnarray}
&\bullet&V_C^+=
\frac{3g_A^4}{16}\,G_1-\frac{3g_A^2\alpha}{2}\,\Bigg\{
4m\,c_1\,\bigg[2\,I_2-I_1-2\alpha(H_1-H_2)\bigg]
\nonumber\\[2mm]&&
+\frac{m\,c_2}{3}\,\alpha\,\Big(3\,H_1-2\,H_3\Big)
-2m\,c_3\,\bigg[I_1-I_3
+\alpha\Big(2\,H_1-2\,H_2-H_3\Big)\bigg]
\Bigg\}
\nonumber\\[2mm]&&
+\frac{3\alpha^2}{2}\,\Bigg[
(4m\,c_1)^2\,H_1+\frac{1}{5}\,(m\,c_2)^2\,\Big(4\,H_2-H_3\Big)
+(2m\,c_3)^2\,\Big(H_1-H_3\Big)
\nonumber\\[2mm]&&
-\frac{16}{3}\,m^2c_1\,c_2\,H_2
-16m^2c_1\,c_3\,\Big(2\,H_2-H_1\Big)
+\frac{4}{3}\,m^2c_2\,c_3\,\Big(2\,H_2-H_3\Big)\Bigg]
+\frac{3g_A^6\mu^2}{16\pi\,f_{\pi}^2}\;\Big(I_1-I_3\Big)
\nonumber\\[2mm]&&
-\frac{3g_A^4\mu^2}{256\pi^2f_{\pi}^2}\bigg\{
\Big[8\,\Big(I_6-I_8\Big)
-7\,\Big(I_5-I_7\Big)\Big]
+4\pi\,\Big[4\,I_3+6\,I_2-7\,I_1\Big]\,
\bigg\}\;,
\end{eqnarray}

\begin{eqnarray}
&\bullet&V_{LS}^+=
\frac{3g_A^4\,\alpha}{8}\,G_2-4g_A^2\,\alpha^2\;m\,c_2\;H_5\;,
\end{eqnarray}

\begin{eqnarray}
&\bullet&V_T^+=
-\frac{g_A^4}{16}\,G_3+\frac{g_A^2\,\alpha^2}{3}\,
(m^2\,\tilde{d}_{14}-m^2\,\tilde{d}_{15})\,\Big(H_3-3\,H_5\Big)
-\frac{g_A^6\,\mu^2}{96\pi^2f_{\pi}^2}\;\Big(H_3-3\,H_5\Big)\;,
\end{eqnarray}

\begin{eqnarray}
&\bullet&V_{SS}^+=
\frac{g_A^4}{8}\,G_4-\frac{2g_A^2\,\alpha^2}{3}\,
(m^2\,\tilde{d}_{14}-m^2\,\tilde{d}_{15})\,H_3
+\frac{g_A^6\,\mu^2}{48\pi^2f_{\pi}^2}\;H_3\;,
\end{eqnarray}

\begin{eqnarray}
&\bullet&V_C^-=
\frac{g_A^4}{8}\,G_5+\frac{g_A^2}{12}\,\Bigg\{\Bigg[
2\Big(5\,H_2-3\,H_1\Big)
-3\,\alpha\,\Big(I_1-I_3\Big)
-3\,\alpha^2\,\Big(2\,H1-2\,H2-H_3\Big)\Bigg]
\nonumber\\[1mm]&&
+\alpha^2\,\Bigg[2m\,c_4\,\Big(5\,H_3-12\,H_1+12\,H_2\Big)
-8(m^2d_1+m^2d_2)\,\Big(5\,H_3+2\,H_2-6\,H_1\Big)
\nonumber\\[1mm]&&
-\frac{4m^2d_3}{5}\,\Big(7\,H_3-8\,H_2\Big)
+32m^2d_5\,\Big(5\,H_2-3\,H_1\Big)\Bigg]\Bigg\}
\nonumber\\[1mm]&&
+\frac{1}{12}\Bigg\{H_2+\alpha^2\,\Bigg[2m\,c_4\,H_3
+8(m^2d_1+m^2d_2)\,\Big(2\,H_2-H_3\Big)
+\frac{12m^2d_3}{5}\,\Big(4\,H_2-H_3\Big)
\nonumber\\[1mm]&&
+32m^2d_5\,H_2\Bigg]\Bigg\}
-\frac{g_A^6\,\mu^2}{288\pi^2f_{\pi}^2}\;
\frac{1}{50}\left(201\,H_3+156\,H_2-300\,H_1\right)
\nonumber\\[1mm]&&
-\frac{g_A^4\,\mu^2}{288\pi^2f_{\pi}^2}\,\frac{1}{50}\,\Big[
1250\,H_8-1500\,H_7+450\,H_6
-346\,H_3+1084\,H_2-300\,H_1
\Big]
\nonumber\\[1mm]&&
-\frac{g_A^2\,\mu^2}{144\pi^2f_{\pi}^2}\,\Bigg\{
5\,H_8-3\,H_7-\frac{61}{20}\,H_3
+\frac{61}{5}\,H_2-3\,H_1
\Bigg\}
\nonumber\\[1mm]&&
-\frac{\mu^2}{288\pi^2f_{\pi}^2}\,\bigg[
H_8
-\frac{1}{10}\Big(14\,H_3-76\,H_2\Big)
\bigg]\,,
\end{eqnarray}

\begin{eqnarray}
&\bullet&V_{LS}^-=
\frac{g_A^4\,\alpha}{16}\,G_6
-\frac{g_A^2\,\alpha}{12}\,\bigg[6\,I_4
+\alpha\,(3-40m\,c_4)\,H_5
+\alpha\,24m\,c_4\,H_4\bigg]
\nonumber\\[1mm]&&
+\frac{\alpha^2}{24}\;(3+16m\,c_4)\,H_5\;,
\end{eqnarray}

\begin{eqnarray}
&\bullet&V_T^-=
\frac{g_A^4\,\alpha}{48}\,G_7
+\frac{g_A^2\,\alpha}{144}\,(1+4m\,c_4)\,
\bigg\{6\,\Big(I_3-3\,I_4\Big)
+\alpha\,\Big[\Big(8\,H_3-12\,H_1+12\,H_2\Big)
\nonumber\\[1mm]&&
-3\,\Big(8\,H_5-3\,H_4\Big)\Big]\bigg\}
-\frac{\alpha^2}{144}\,(1+4m\,c_4)^2\,\Big(H_3-3\,H_5\Big)
-\frac{g_A^6\,\mu^2}{96\pi\,f_\pi^2}\,\Big(I_3-3\,I_4\Big)
\nonumber\\[1mm]&&
+\frac{g_A^4\,\mu^2}{384\pi^2f_\pi^2}\bigg[
\Big(I_8-3\,I_9\Big)
-2\pi\,\Big(I_3-3\,I_4\Big)\bigg]\;,
\end{eqnarray}

\begin{eqnarray}
&\bullet&V_{SS}^-=
-\frac{g_A^4\,\alpha}{24}\,G_8
-\frac{g_A^2\,\alpha}{72}\,(1+4m\,c_4)\,\bigg[6\,I_3
+\alpha\,\Big(8\,H_3-12\,H_1+12\,H_2\Big)\bigg]
\nonumber\\[1mm]&&
+\frac{\alpha^2}{72}\,(1+4m\,c_4)^2\,H_3
+\frac{g_A^6\,\mu^2}{48\pi\,f_\pi^2}\,I_3
-\frac{g_A^4\,\mu^2}{192\pi^2f_\pi^2}\,\Big(I_8
-2\pi\,I_3\Big)\;.
\end{eqnarray}

\newpage

\begin{table}[htb]
\begin{tabular} {|l||c|c|c|c|c|c|}
\hline
$\g_i^n$ & $-1/2$ & $-3/2$ & $-5/2$ & $-7/2$ & $-9/2$ & $-11/2$ \\ \hline 

$H_1$ & -1 & -3/16 & 15/512 & -105/8192 & 0.0069211 & -0.002031054 \\ \hline 
$H_2$ & - & 3/2	& 45/32 & 315/1024 & -0.050879 & 0.0105639 \\ \hline 
$H_3$ & - & 6 & 165/8 & 8715/256 & 27.45483 & 5.43256 \\ \hline 
$H_4$ & - & 2 & 23/8 & 153/256 & -0.0723934 & - \\ \hline 
$H_5$ & - & - & -3 & -129/16 & -3555/512 & -1.33605 \\ \hline 
$H_6$ & -3.89861 & 4.23305 & -0.833136 & - & - & - \\ \hline 
$H_7$ & - & 5.78893 & -7.63019 & -2.69576 & - & - \\ \hline 
$H_8$ & - & - & -14.3654 & 14.6375 & 39.3909 & 18.8729 

\end{tabular}
\label{T3}
\end{table}


\begin{table}[htb]
\begin{tabular} {|l||c|c|c|c|c|c|c|c|}
\hline
$\g_i^n$ & 1 & $0$ & $-1$ & $-2$ & $-3$ & $-4$ & $-5$ & $-6$ \\ \hline 

$G_1$ & - & 2.83823 & -7.200711 & 38.9637 & -55.5164 & 47.2443 
& -16.2395 & - \\ \hline 
$G_2$ & - & - & -6.12315 & -28.1422 & -30.2813 & 0.023458 & -15.8996
& 7.18869 \\ \hline 
$G_3$ & - & 0.5579 & 17.1039 & 16.8038 & 9.94755 & 3.40171 & -2.7544 
& - \\ \hline 
$G_4$ & - & 0.569624 & 15.9429 & -4.26031 & 15.6445 & -5.06641 & - 
& - \\ \hline 
$G_5$ & -0.217221 & -9.98415 & -4.662 & -36.9761 & 13.4087 & -6.21047 
& - & - \\ \hline 
$G_6$ & - & - & 7.90985 & 55.9568 & 86.3242 & 66.9540 & -29.5680 
& 11.8985 \\ \hline 
$G_7$ & - & 1.69219 & 25.5612 & 6.53589 & 160.459 & -169.567 & 120.612 
& -36.7881 \\ \hline 
$G_8$ & - & 1.7661 & 21.2122 & -9.87710 & 116.454 & -144.344 & 103.063 
& -30.9265 

\end{tabular}
\label{T1}
\end{table}


\begin{table}[htb]
\begin{tabular} {|l||c|c|c|c|c|c|c|}
\hline
$\g_i^n$ & $2$ & $1$ & $0$ & $-1$ & $-2$ & $-3$ & $-4$ \\ \hline 

$I_1$ & 0.000483761 & -0.0226386 & 1.53346 & 0.0595627 & -0.0913580 
& 0.0291743 & - \\ \hline 
$I_2$ & - & -0.000381934 & 0.0158372 & -1.63009 & -0.660019 
& 0.0532419 & - \\ \hline 
$I_3$ & - & - & 0.242214 & -8.87827 & -6.47733 & -30.5206 & - \\ \hline 
$I_4$ & - & - & - & -0.00147946 & 2.99191 & 6.86185 & 1.82098 

\end{tabular}
\label{T4}
\end{table}


\begin{table}[htb]
\begin{tabular} {|l||c|c|c|c|c|c|c|}
\hline
$\g_i^n$ & $3/2$ & $1/2$ & $-1/2$ & $-3/2$ & $-5/2$ & $-7/2$ 
& $-9/2$ \\ \hline 
$I_5$ & 0.168731 & -6.00262 & -2.18325 & 1.79108 & - & - & - \\ \hline 
$I_6$ & - & -0.129344 & 5.46315 & -2.54740 & 0.240122 & - & - \\ \hline 
$I_7$ & - & -0.464737 & 20.2994 & 31.4762 & -26.1396 & - & - \\ \hline 
$I_8$ & - & - & - & -28.6452 & -101.827 & 28.3771 & 99.2609 \\ \hline 
$I_9$ & - & - & - & -0.454235 & 18.7535 & 24.4758 & -31.2207 

\end{tabular}
\label{T5}
\end{table}

The parametrized profile functions given 
above depend explicitly on four well known quantities, namely $m$, 
$\m$, $g_A$, $f_\pi$, and on the less known LECs $c_i$ and $d_i$. 
Therefore, the 
latter may be extracted from fits to data. When doing this, however, 
one has to bear in mind that, as discused in Ref. {\cite{HRR}}, the 
influence of the LECs over the profile functions is rather uneven.
Indeed, their influence over $V_C^+$, $V_{LS}^-$, $V_T^-$, and 
$V_{SS}^-$ is rahter strong, but barely perceptible in $V_C^-$, 
$V_{LS}^+$, $V_T^+$ and $V_{SS}^+$. 

\newpage

\section*{acknowledgments}

The work of R.~H. was supported by DOE Contract No.DE-AC05-84ER40150 
under which SURA operates the Thomas Jefferson National Accelerator 
Facility, and M.~R.~R. and C.~A.~dR. were supported by FAPESP 
(Funda\c c\~ao de Amparo \`a Pesquisa do Estado de S\~ao Paulo). 


\appendix\section{theoretical potential}

The $O(q^4)$ relativistic expansion of the TPEP produced in 
Refs.~\cite{HR,HRR} was based on the evaluation of three familes of 
diagrams\footnote{Please see section II of Ref. \cite{HRR}, for a 
detailed discussion of the meaning and dynamical contents of these 
families of diagrams, which are given in its Fig.2.}. The first of 
them involves only  pion and nucleon degrees of freedom into single 
loops and corresponds to the minimal realization of chiral symmetry 
\cite{SP}. It includes the subtraction of the iterated OPEP and yields 
the terms in the profile functions given below which are proportional 
to just $g_A^4/f_\p^4$, $g_A^2/f_\p^4$ or $1/f_\p^4$. Terms 
proportional to $1/f_\p^6$, on the other hand, come from two-loop 
processes, either in the form of $t$-channel contributions from the 
second family or $s$ and $u$-channel terms embodied in the subthreshold 
coefficients of the third family. Finally, the third group of diagrams 
includes chiral corrections associated with other degrees of freedom, 
hidden within the LECs $c_i$ and $d_i$, and gives rise to contributions 
which are proportional to either $(LEC)/f_\p^4$ or  $(LEC)^2/f_\p^4$. 
In Ref. \cite{HRR} we have expressed this last class of results in 
terms of the $\pi N$ subthreshold coefficients. As these can be easily 
translated into LECs, in the present work we write the potential in 
terms of these constants, which appear directly into the effective 
lagrangians. The following expressions correspond to the updated 
version as described in Ref.~\cite{R04}. 

The radial components of the potential are expressed in terms of the 
following profile functions\footnote{Please see Eqs. (3.4)-(3.8) 
of Ref.~\cite{HRR}. Note that in Eq. (3.8) a multiplication factor of 
$\mu^3$ is missing.} 

\begin{eqnarray}
V_{C}^{\pm}(r) &=& \tau^{\pm}\;U_{C}^{\pm}(x)\,,
\label{3.4}\\[2mm]
V_{LS}^{\pm}(r) &=& \tau^{\pm}\;\frac{\mu^2}{m^2}\,\frac{1}{x}\,\frac{d}{dx}\,
U_{LS}^{\pm}(x)\,,
\label{3.5}\\[2mm]
V_{T}^{\pm}(r) &=& \tau^{\pm}\;\frac{\mu^2}{m^2}\left[\frac{d^2}{d x^2}
-\frac{1}{x}\,\frac{d}{dx}\right]\,U_{T}^{\pm}(x)\,,
\label{3.6}\\[2mm]
V_{SS}^{\pm}(r) &=& -\tau^{\pm}\;\frac{\mu^2}{m^2}\left[\frac{d^2}{dx^2}
+\frac{2}{x}\,\frac{d}{dx}\right]\,U_{SS}^{\pm}(x)\,,
\label{3.7}
\end{eqnarray}

\noindent where $\tau^+=3$ and $\tau^-=2$. 

Defining $\hat{t}$ as the Laplacian operator acting on the variable 
$x=\mu r$ we write our profile functions as 

\begin{eqnarray}
&&-U_C^+=
\frac{3g_A^4\mu^5}{256\pi^2f_{\pi}^4}\,
\bigg\{
\left[1-\left(1-\alpha^2/4\right)\hat{t}
+\hat{t}^2/4\right]\,S_\times
-\left[1-\left(1+\alpha^2/4\right)\hat{t}
+\hat{t}^2/4\right]\,S_b
\nonumber\\[2mm]&&
-\alpha\,(1-\hat{t}/2)\left[2S_a+\hat{t}\,S_t\right]
+\alpha^2\,\hat{t}^2\,S_\ell\bigg\}
-\frac{3g_A^2\mu^5\,\alpha}{32\pi^2f_{\pi}^4}\,\Bigg\{
4m\,c_1\,\bigg[\Big[2(1-\hat{t}/4)-1\Big]\,S_t
-\frac{\alpha\,\hat{t}}{2}\,S_\ell\bigg]
\nonumber\\[2mm]&&
+\frac{m\,c_2}{3}\,\alpha\left[3-2(1-\hat{t}/4)\,\hat{t}\,\right]S_\ell
-2m\,c_3\,\bigg[\Big[1-(1-\hat{t}/4)\,\hat{t}\,\Big]\,S_t
+\alpha\left[\hat{t}/2-(1-\hat{t}/4)\,\hat{t}\,\right]S_\ell\bigg]
\Bigg\}
\nonumber\\[2mm]&&
+\frac{3\mu^5\,\alpha^2}{32\pi^2f_{\pi}^4}\,\Bigg\{
(4m\,c_1)^2\,S_\ell
+\frac{1}{5}\,(m\,c_2)^2\,
\Big[4(1-\hat{t}/4)-(1-\hat{t}/4)\,\hat{t}\,\Big]\,S_\ell
\nonumber\\[2mm]&&
+(2m\,c_3)^2\,\left[1-(1-\hat{t}/4)\,\hat{t}\,\right]S_\ell
-\frac{16}{3}\,m^2c_1\,c_2\,(1-\hat{t}/4)\,S_\ell
-16m^2c_1\,c_3\,\left[2(1-\hat{t}/4)-1\right]S_\ell
\nonumber\\[2mm]&&
+\frac{4}{3}\,m^2c_2\,c_3\,
\Big[2(1-\hat{t}/4)-(1-\hat{t}/4)\,\hat{t}\Big]\,S_\ell\Bigg\}
+\frac{3g_A^6\mu^7}{256\pi^3f_{\pi}^6}\;
\Big[1-(1-\hat{t}/4)\,\hat{t}\,\Big]\,S_t
\nonumber\\[2mm]&&
-\frac{3g_A^4\mu^7}{4096\pi^4f_{\pi}^6}\bigg\{
\Big\{8(1-\hat{t}/4)-7
-\Big[8(1-\hat{t}/4)^2-7(1-\hat{t}/4)\Big]\,\hat{t}\,\Big\}\,S_{tt}
\nonumber\\[2mm]&&
+4\pi\,\Big[4(1-\hat{t}/4)\,\hat{t}+6(1-\hat{t}/4)-7\Big]\,S_t\,
\bigg\}\;,
\end{eqnarray}

\begin{eqnarray}
&&-U_{LS}^+=
\frac{3m\,g_A^4\mu^4}{128\pi^2f_{\pi}^4}\,
\bigg\{ (1-\hat{t}/2)\,(\tilde S_b-S_t)
-(3/2-5\,\hat{t}/8)\,S_a
+\frac{\alpha}{4}\,(1\!+\!2\hat{t}\!-\!\hat{t}^2/2)\,(S_\times\!+\!S_b)
\nonumber\\[2mm]&&
+2\alpha\,\hat{t}\,S_\ell\bigg\}
-\frac{g_A^2\mu^5}{4\pi^2f_{\pi}^4}\;m\,c_2\;(1\!-\!\hat{t}/4)\,S_\ell\;,
\end{eqnarray}

\begin{eqnarray}
&&-U_T^+=-U_{SS}^+/2=
-\frac{m^2g_A^4\mu^3}{256\pi^2f_{\pi}^4}\,
\bigg\{(1-\hat{t}/4)\,S_b
+\frac{\alpha}{2}\left[(1-\hat{t}/2)\,(S_t-\tilde S_b)
+(1-\hat{t}/4)\,S_a\right]
\nonumber\\[2mm]&&
+\left[\left(1-\alpha^2/4\right)-\left(1-\alpha^2\right)
\hat{t}/4-\alpha^2\,\hat{t}^2/16\right]S_\times\bigg\}
+\frac{g_A^2\mu^5}{48\pi^2f_{\pi}^4}\,
(m^2\,\tilde{d}_{14}-m^2\,\tilde{d}_{15})\,(1-\hat{t}/4)\,S_\ell
\nonumber\\[2mm]&&
-\frac{m^2\,g_A^6\mu^5}{1536\pi^4f_{\pi}^6}\;(1-\hat{t}/4)\,S_\ell\;,
\end{eqnarray}

\begin{eqnarray}
&&-U_C^-=
\frac{g_A^4\mu^5}{128\pi^2f_{\pi}^4}
\bigg\{
\left[1-\left(1-\alpha^2/4\right)\hat{t}
+\left(1-\alpha^2\right)\hat{t}^2/4
+\alpha^2\,\hat{t}^3/16\right]S_\times
\nonumber\\[2mm]&&
+\left[1-\left(1+\alpha^2/4\right)\hat{t}
+\left(1+\alpha^2\right)\hat{t}^2/4
-\alpha^2\,\hat{t}^3/16\right]S_b
\nonumber\\[2mm]&&
+\alpha\Big[(2-3\hat{t}+\hat{t}^2)\,S_t+(2-\hat{t})S_a\Big]
-\left[
10/3-\left(11/6-\alpha^2\right)\hat{t}-\alpha^2\,\hat{t}^2/2\right]S_\ell
\bigg\}
\nonumber\\[2mm]&&
+\frac{g_A^2\mu^5}{192\pi^2f_{\pi}^4}\,\Bigg\{
2\left[5(1-\hat{t}/4)-3\right]\,S_\ell
-3\,\alpha\left[1-(1-\hat{t}/4)\,\hat{t}\right]S_t
-3\,\alpha^2\,\Big[\hat{t}/2-(1-\hat{t}/4)\,\hat{t}\Big]\,S_\ell
\nonumber\\[2mm]&&
+\alpha^2\,\bigg[
2m\,c_4\,\left[5(1-\hat{t}/4)\,\hat{t}-3\,\hat{t}\right]\,S_\ell
-8(m^2d_1+m^2d_2)\,\left[5(1-\hat{t}/4)\,\hat{t}+2(1-\hat{t}/4)-6\right]\,S_\ell
\nonumber\\[2mm]&&
-\frac{4m^2d_3}{5}\,\left[7(1-\hat{t}/4)\,\hat{t}-8(1-\hat{t}/4)\right]\,S_\ell
+32m^2d_5\,\left[5(1-\hat{t}/4)-3\right]\,S_\ell\bigg]\Bigg\}
\nonumber\\[2mm]&&
+\frac{\mu^5}{192\pi^2f_{\pi}^4}\Bigg\{\,(1-\hat{t}/4)\,S_\ell
+\alpha^2\bigg[2m\,c_4\,\Big[(1-\hat{t}/4)\,\hat{t}\Big]\,S_\ell
\nonumber\\[2mm]&&
+8(m^2d_1+m^2d_2)\,\Big[2(1-\hat{t}/4)-(1-\hat{t}/4)\,\hat{t}\Big]\,S_\ell
+\frac{12m^2d_3}{5}\,\Big[4(1-\hat{t}/4)-(1-\hat{t}/4)\,\hat{t}\Big]\,S_\ell
\nonumber\\[2mm]&&
+32m^2d_5\,(1-\hat{t}/4)\,S_\ell\bigg]\Bigg\}
-\frac{g_A^6\mu^7}{4608\pi^4f_{\pi}^6}\;
\frac{1}{50}\left[201(1-\hat{t}/4)\,\hat{t}+156(1-\hat{t}/4)-300\right]\,S_\ell
\nonumber\\[2mm]&&
-\frac{g_A^4\mu^7}{4608\pi^4f_{\pi}^6}\;\frac{1}{50}\Bigg\{
\Big[1250(1-\hat{t}/4)^2-1500(1-\hat{t}/4)+450\Big]\,S_{\ell\ell}
\nonumber\\[2mm]&&
+\left[-346(1-\hat{t}/4)\,\hat{t}+1084(1-\hat{t}/4)-300\right]\,S_\ell\,
\Bigg\}
\nonumber\\[2mm]&&
-\frac{g_A^2\mu^7}{2304\pi^4f_{\pi}^6}\,\Bigg\{
\left[5(1-\hat{t}/4)^2-3(1-\hat{t}/4)\right]\,S_{\ell\ell}
+\left[-\frac{61}{20}\,(1-\hat{t}/4)\,\hat{t}+\frac{61}{5}\,(1-\hat{t}/4)
-3\right]\,S_\ell\Bigg\}
\nonumber\\[2mm]&&
-\frac{\mu^7}{4608\pi^4f_{\pi}^6}\,\Bigg\{
(1-\hat{t}/4)^2\,S_{\ell\ell}
-\frac{1}{10}\left[14(1-\hat{t}/4)\,\hat{t}
-76(1-\hat{t}/4)\right]\,S_\ell\Bigg\}\,,
\end{eqnarray}

\begin{eqnarray}
&&-U_{LS}^-=
\frac{m\,g_A^4\mu^4}{256\pi^2f_\pi^4}\,
\bigg\{
(6-5\,\hat{t}/2)\,S_a-(4-2\,\hat{t})\,\tilde S_b+4\,S_t
\nonumber\\[2mm]&&
+\alpha\,\Big[2\,(1-\hat{t}/4)\,S_\ell
+(1-\hat{t}/2)^2\,(S_\times-S_b)\Big]\bigg\}
\nonumber\\[2mm]&&
-\frac{m\,g_A^2\mu^4}{192\pi^2f_\pi^4}\,\bigg[6\,(1-\hat{t}/4)\,S_t
+\alpha\,(3-40m\,c_4)\,(1-\hat{t}/4)\,S_\ell
+\alpha\,24m\,c_4\,S_\ell\bigg]
\nonumber\\[2mm]&&
+\frac{\mu^5}{384\pi^2f_\pi^4}\;(3+16m\,c_4)\,(1-\hat{t}/4)\,S_\ell\;,
\end{eqnarray}

\begin{eqnarray}
&&-U_T^-=-U_{SS}^-/2=
\frac{m\,g_A^4\mu^4}{768\pi^2f_\pi^4}\,
\bigg[ (1\!-\!\hat{t}/2)\,\tilde S_b
+(1\!-\!\hat{t}/4)\,(S_a-2\,S_t)
-\frac{\alpha}{12}\,(16-7\,\hat{t})\,S_\ell\bigg]
\nonumber\\[2mm]&&
+\frac{m\,g_A^2\mu^4}{2304\pi^2f_{\pi}^4}\,(1+4m\,c_4)\,\bigg\{
6\,(1-\hat{t}/4)\,S_t+\alpha\left[8\,(1-\hat{t}/4)-3\right]S_\ell\bigg\}
\nonumber\\[2mm]&&
-\frac{\mu^5}{2304\pi^2f_\pi^4}\,(1+4m\,c_4)^2\,(1-\hat{t}/4)\,S_\ell
-\frac{m^2g_A^6\mu^5}{1536\pi^3f_\pi^6}\,(1-\hat{t}/4)\,S_t
\nonumber\\[2mm]&&
+\frac{m^2g_A^4\mu^5}{6144\pi^4f_\pi^6}\bigg[
(1-\hat{t}/4)^2\,S_{tt}-2\pi\,(1-\hat{t}/4)\,S_t\bigg]\;.
\end{eqnarray}

The dimensionless functions $S_i(x)$ carry the spatial dependence of 
the potential and are given by Eqs. (3.16)-(3.23) of Ref.~\cite{HRR}.


\end{document}